\documentclass[aps,prl,reprint,preprintnumbers,superscriptaddress,amsmath,amssymb,bibnotes,longbibliography]{revtex4-1}

\usepackage{graphicx}
\usepackage{bm}
\usepackage[colorlinks,linkcolor=blue,anchorcolor=blue,citecolor=blue,urlcolor=blue,filecolor=blue,menucolor=blue,runcolor=blue]{hyperref}

\graphicspath{{figures/}}

\begin{document}


\title{Strong enhancement of magnetic coercivity induced by uniaxial stress}

\author{Bin Shen}
\email{bin.shen@physik.uni-augsburg.de}
\affiliation  {Experimental Physics VI, Center for Electronic Correlations and Magnetism, University of Augsburg, 86159 Augsburg, Germany}

\author{Franziska Breitner}
\affiliation  {Experimental Physics VI, Center for Electronic Correlations and Magnetism, University of Augsburg, 86159 Augsburg, Germany}

\author{Philipp Gegenwart}
\affiliation{Experimental Physics VI, Center for Electronic Correlations and Magnetism, University of Augsburg, 86159 Augsburg, Germany}

\author{Anton Jesche}
\affiliation  {Experimental Physics VI, Center for Electronic Correlations and Magnetism, University of Augsburg, 86159 Augsburg, Germany}

\date{\today}

\begin{abstract}   
The performance of permanent magnets is intricately tied to their magnetic hysteresis loop. In this study, we investigate the heavy-fermion ferromagnet CeAgSb$_2$ through magnetization measurements under uniaxial stress. We observe a 2400~\% increase in magnetic coercivity with just a modest stress of approximately 1~kbar. This effect persists even after pressure release, attributable to stress-induced defects that efficiently pin domain walls. Other magnetic properties such as ordering temperature and saturation moment exhibit only weak pressure dependencies and display full reversibility. Our findings offer a promising route for increasing coercive field strength and enhancing the energy product in ferromagnetic materials and are potentially applicable to a broad spectrum of commercial or emerging magnetic applications.
\end{abstract}

\maketitle


Uniaxial stress $p$ or its response -- strain (characterized by the relative length change $\Delta L/L$) -- provides unique access to directional tuning of the lattice in a material. Such effects can drive modifications of phases or functionality of the material, which has been widely applied in various systems, such as unconventional superconductors \cite{14ScienceCli, 17ScienceAle, 18ScienceKim}, perovskite manganites \cite{20ScienceSeu}, topological semimetals \cite{19SAJos}, and frustrated magnets \cite{17PRBKuc}.   However, stress effects in permanent magnets remain largely unexplored. 

Permanent magnetic materials are widely used in various forms of energy conversion applications, such as motors and generators, and their demand is growing especially in the field of renewable energy generation \cite{2011Gut}. A fundamental attribute of a permanent magnet is its magnetic hysteresis loop. In order to achieve high energy product, a high coercivity $H_{\rm{C}}$ and a large remnant moment $M_{\rm{r}}$ are of paramount importance \cite{2011Gut}. Since the synthesis of early generation magnet (high-carbon steel) with $H_{\rm{C}}$ around 100~Oe, a plethora of materials with higher $H_{\rm{C}}$ have been manufactured \cite{2012Gary, 2017Kris}. The development of Sm-Co and, especially, Nd-Fe-B magnets have inaugurated a golden age for application of magnets thanks to the record-high values of $H_{\rm{C}}$ (10--30~kOe) and energy product \cite{2012Gary, 1984Sag, 1984Cro}.  

By nature, coercivity is an extrinsic property of a magnet. It is bounded by the anisotropy field $H_{\rm{A}}$ as proposed in the Stoner-Wohlfarth model \cite{1948Stoner}. Therefore, the incorporation of rare earth elements is usually a shared characteristic in the production of permanent magnets because of their strong spin-orbit-coupling which is essential for a large magneto-crystalline anisotropy. However, the theoretical limit of coercivity has never been achieved experimentally \cite{2000Gut}. Coercivity, in most cases, only reaches 20--30~\% of $H_{\rm{A}}$ \cite{2000Gut}, also known as Brown's paradox \cite{1945Bro}. Such practical reduction of the coercivity is due to collective reversal of magnetic moments via propagation of domain walls. Hence, in principle, impeding the nucleation or pinning domain walls through introducing certain types of defects will enhance $H_{\rm{C}}$ \cite{1981Liv}.

 \begin{figure*}
	\includegraphics[angle=0,width=0.98\textwidth]{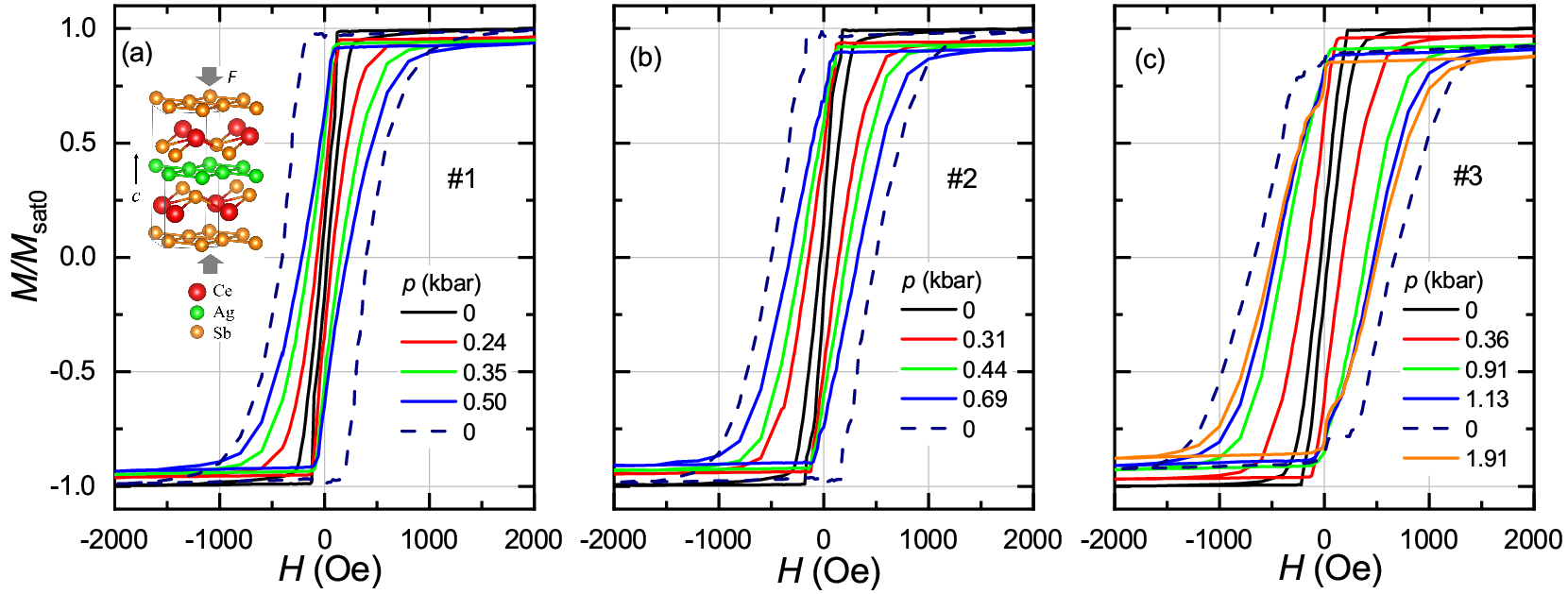}
	\vspace{-12pt} \caption{\label{Figure1} Normalized magnetization $M/M_{\rm{sat0}}$ as a function of the magnetic field $H$ with field applied along the $c$ axis at 5~K at various stresses for (a) sample \#1, (b) \#2, and (c) \#3. Dashed lines stand for the measurement after complete decompression. Inset of (a) shows the crystal structure of CeAgSb$_2$. Solid lines indicate one unit cell. Stress was applied along the $c$ axis. For sample \#3, the stress of 1.91~kbar was applied after total decompression.}
	\vspace{-12pt}
\end{figure*}

Various approaches have been implemented to enhance $H_{\rm{C}}$ for magnets. They are usually realized by composition-tuning \cite{ai, ad, aq, aj, am, an, ao, ap}, modifying different sintering and/or annealing temperatures \cite{aa, ak, al}, varying particle sizes \cite{ah, ab, ac, ae, af}, and strain-tuning using different substrates for thin-films \cite{ar}. The basic aim of these approaches is to induce structural, chemical, or magnetic imperfections in the magnet. However, the underlying mechanism responsible for the increase in coercivity by each method arguably remains not well understood. Often a combination of different methods will be employed simultaneously to a magnet to optimize its functionality while producing high coercivity. For instance, a sintered Dy-free Nd-Fe-B magnet with a $H_{\rm{C}}$ of 6.3~kOe needs to go through the so-called grain boundary diffusion (GBD) process combined with annealing to reach final coercivity of 20.7~kOe \cite{ag}.  Nonetheless, it seems that the quest for higher coercivity has reached a technological bottleneck.

In this letter, we present findings from magnetization measurements conducted on the Kondo lattice CeAgSb$_2$, revealing a previously unrecognized and highly efficient method for enhancing the coercivity: uniaxial stress. Surprisingly, even a modest stress of approximately 1~kbar causes a remarkable increase in coercive field by approximately 2400~\%. This enhancement is likely attributed to stress-induced defects, effectively transitioning CeAgSb$_2$ from a soft magnet to a hard one. This discovery was unexpected, as our original intent was to explore stress-induced (quantum) phase transitions in CeAgSb$_2$.

The Kondo lattice compound CeAgSb$_2$ undergoes a ferromagnetic transition at $T_{\rm{C}}$ = 9.7~K \cite{95JSSCSol, 03PRBTak}. Crystal electric field in the compound renders the $ab$ plane the easy plane. However, the ordered moment of about 0.4~$\mu_{\rm{B}}$ points along the $c$ axis \cite{03PRBAra}, which is a generic property of ferromagnetic Kondo lattice systems \cite{19PRBHaf}. The ferromagnetic order can be suppressed by magnetic field applied in the $ab$ plane \cite{13PSSLog, 13PSSZou, 18JPSJKaw, 21PRBNik}, or by hydrostatic pressure \cite{03JPCMMih, 03PRBSid, 13PSSLog},  leading to exotic quantum phase transitions. Shubnikov-de Haas oscillation \cite{07PRBpRO} and quantum oscillation in the thermopower \cite{11JPCMMun} were observed in CeAgSb$_2$ at around 2~K. For the latter case, the thermopower starts to oscillate at a relatively low field of 1.5~T \cite{11JPCMMun}. Using ultrafast pump probe, a possible orbital crossover was proposed in CeAgSb$_2$ \cite{2019Che}. Recently, a surprising topological magnetic hysteresis with rectangular tubular pattern of the magnetic domains was  uncovered in thick CeAgSb$_2$ single crystals \cite{22JPCMRus}, which is usually observed in thin films, making CeAgSb$_2$ an even more intriguing ferromagnet.

Single crystals of CeAgSb$_2$ were grown using an Sb-rich flux \cite{03PRBTak}. Magnetization with magnetic field applied along the crystallographic $c$-axis under uniaxial stress was measured in a Quantum Design MPMS system. Stress was exerted along the crystallographic $c$-axis, with magnetization subsequently measured along this same direction [Fig.~\ref{Figure1} (a)]. The details of sample preparation and magnetization measurement are described in the Supplementary Materials \cite{CeAgSbSI}. Electrical transport measurements are done in a Quantum Design Dynacool system using standard four-point probe. Electron micrographs were recorded using a ZEISS Merlin system.   

Isothermal magnetization loops $M(H)$ collected at $T$ = 5 K are shown in [Figs.~\ref{Figure1} (a-c)] for three different samples. The obtained $M(H)$ values were normalized by the corresponding saturation magnetization $M_{\rm{sat0}}$ at zero stress that amounts to 0.34~$\mu_{\rm{B}}$, 0.40~$\mu_{\rm{B}}$, and 0.35~$\mu_{\rm{B}}$ for sample \#1, \#2, and \#3, respectively. Given that the small sample size leads to a relative mass error of roughly 10~\%, the values are considered to be in reasonable agreement with the literature value of 0.4~$\mu_{\rm{B}}$ \cite{03PRBTak, MYERS199927, 22JPCMRus}. At zero stress, $H_{\rm{C}}$ of as-grown CeAgSb$_2$ at 5~K is around 30~Oe, similar to the value reported previously \cite{22JPCMRus}.  Along with a small  $M_{\rm{r}}$ of around 0.05~$\mu_{\rm{B}}$ per Ce, the $M(H)$ hysteresis loop reveals that CeAgSb$_2$ behaves like a soft magnet. 

With the application of uniaxial stress, the hysteresis loop expands significantly as $H_{\rm{C}}$ and $M_{\rm{r}}$ both increases, displaying the box-like shape of a hard magnet. In order to release the pressure and retrieve the sample, a slightly larger force needs to be applied to our stress cell. Subsequently, magnetization after complete decompression was also measured, as indicated by dashed lines in Figs.~\ref{Figure1} (a-c).  Intriguingly, $H_{\rm{C}}$ measured after total decompression shows further enhancement compared to the previous value collected under stress. Moreover, $M_{\rm{r}}$ draws near to the saturation moment after decompression. These results reveal irreversible effects to the hysteresis loop -- that is defects induced by stress.         

\begin{figure}
	
	\includegraphics[angle=0,width=0.32\textwidth]{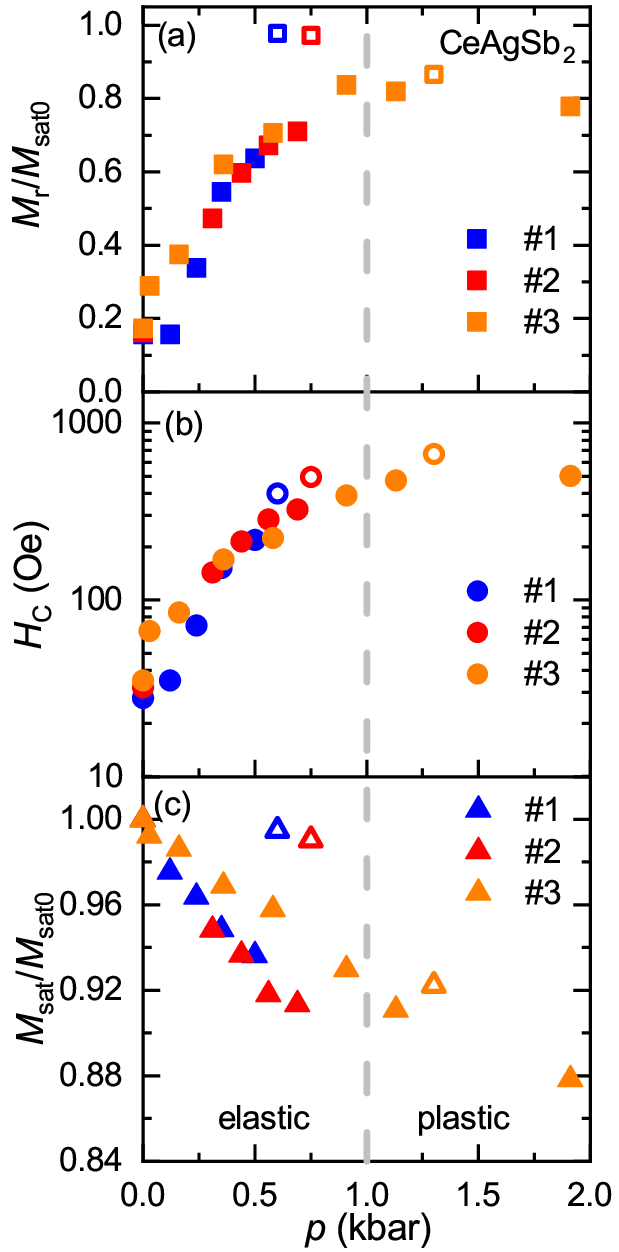}
	\vspace{-12pt} \caption{\label{PD} Variations of (a) normalized remnant moment $M_{\rm{r}}/M_{\rm{sat0}}$ with respect to $M_{\rm{sat0}}
		$, (b) coercivity $H_{\rm{C}}$ in a semi-logarithmic representation, and (c) normalized saturation moment $M_{\rm{sat}}$/$M_{\rm{sat0}}$, as a function of the stress of CeAgSb$_2$ at 5~K. At small stress, sample can be considered in the elastic deformation region. For stress in excess of around 1~kbar, sample enters into the plastic deformation region. Two regions are roughly demarcated by the dashed line. The open symbols were collected after total decompression from the pressures indicated in the phase diagram.}
	\vspace{-12pt}
\end{figure}

Based on 4-quadrant magnetic hysteresis loops, we plot the stress evolution of normalized remnant moment $M_{\rm{r}}/M_{\rm{sat0}}$ and coercive field $H_{\rm{C}}$ in Fig.~\ref{PD}. $M_{\rm{r}}$ increases with increasing stress in a linear fashion up to $p$ $\approx$ 0.8~kbar where it approches the saturation magnetization for samples \#1 and \#2. $H_{\rm{C}}$ increases logarithmically upon compression with a change of slope at $p$ $\approx$ 0.8~kbar reaching values of 400~Oe. The highest $H_{\rm{C}}$ of 670~Oe was recorded after total decompression of sample \#3. 

On the other hand, $M_{\rm{sat}}$ is linearly reduced by around 10~\% upon a stress of 1~kbar, as shown in Figs.~\ref{Figure1} (a-c) and Fig.~\ref{PD} (c). In contrast to $H_{\rm{C}}$ and $M_{\rm{r}}$, $M_{\rm{sat}}$ measured after complete decompression almost exactly returns to its original value for both sample \#1 and \#2. $M_{\rm{sat}}$ of sample \#3 measured after decompression from a higher stress of 1.13~kbar deviates from its original value by about 8~\%. When subsequently subject to a much larger stress of 1.91~kbar, both $M_{\rm{sat}}$ and $H_{\rm{C}}$ of sample \#3 further decrease by 4~\% and 166~Oe, respectively. The coercivity of coarsely (finely) ground samples was also measured (Fig.~S~1 \cite{CeAgSbSI}) and shows a lower value of 200(450)~Oe compared to that measured at 1.91~kbar.

The non-monotonic behavior of $H_{\rm{C}}$ above a certain stress together with the irreversibility of $M_{\rm{sat}}$ indicates the presence of two regions in Fig.~\ref{PD}: elastic and plastic. At low stress (below around 1~kbar), response of the deformation of the sample to stress is basically still elastic. While for larger stress, plastic deformation of the sample begins to play a role as evidenced by the irreversibility of $M_{\rm{sat}}$, which may also account for the decrease of $H_{\rm{C}}$ at highest measured stress of 1.91~kbar.

\begin{figure}
	\includegraphics[angle=0,width=0.48\textwidth]{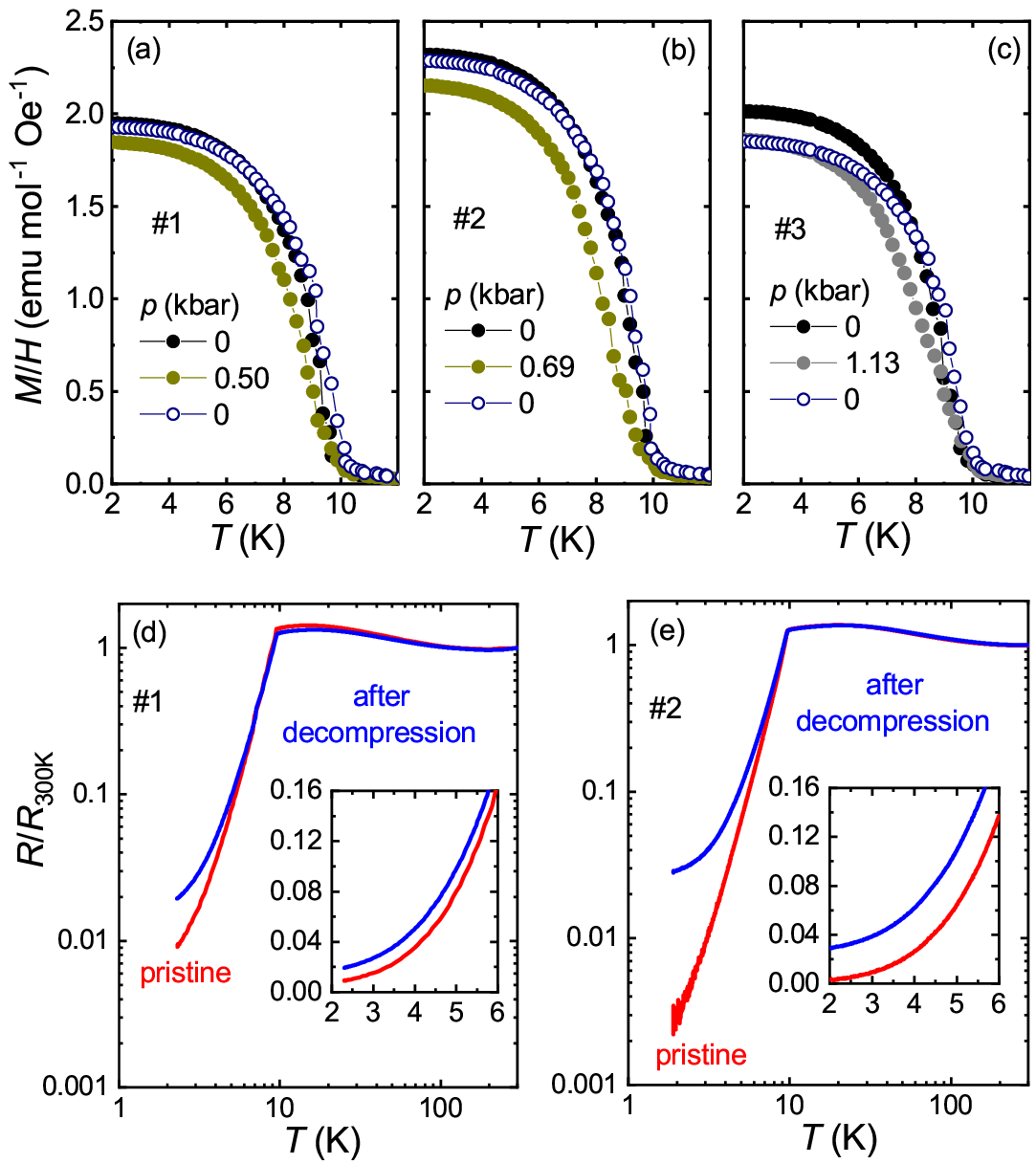}
	\vspace{-12pt} \caption{\label{MT} Magnetization measured at 0.1~T as a function of the temperature $M(T)$ at selective stresses for sample (a) \#1, (b) \#2, and (c) \#3, respectively. The open symbols were collected after decompression. Normalized resistance, with a double-logarithmic representation, measured from pristine sample and from the same one after complete decompression of sample (d) \#1 and (e) \#2, respectively. The insets of (d) and (e) show the low temperature part.}
	\vspace{-12pt}
\end{figure}

To gain further insight into the effects of stress to the sample, we have measured magnetization as a function of the temperature $M(T)$ under various stresses at a magnetic field of 0.1~T, as shown in  Figs.~\ref{MT} (a-c) and Fig.~S~2 \cite{CeAgSbSI}. It can be observed that the ferromagnetic ordering temperature $T_{\rm{C}}$ remains almost unchanged under stress, and that the magnetization shows a slight decrease in the ferromagnetic state in accordance with the small reduction observed in $M(H)$ [Figs.~\ref{Figure1} (a-c)]. After decompression, $M(T)$ of sample \#1 and \#2 almost perfectly overlaps with the uncompressed curve, respectively. Accordingly, pressure seems to have no significant effect on the magnetic exchange. In contrast, $M(T)$ of sample \#3 measured after decompression from a higher stress shows visible deviation from that at ambient condition, further indicating that sample is in the plastic region at stress above 1~kbar.   

Figures.~\ref{MT} (d) and (e) show normalized electrical resistance of the pristine sample and of the same one after total decompression of sample \#1 and \#2, respectively. A simple inspection reveals a practically unchanged behavior in the paramagnetic state and same $T_{\rm{C}}$ before compression and after decompression for both samples. Below $T_{\rm{C}}$, the two resistance curves begin to deviate, implying the modified scattering to the electrons. For sample \#1, the residual resistance ratio (RRR, defined as $R_{300\,{\rm{K}}}$/$R_{2\,{\rm{K}}}$, which can be taken as an indicator for sample quality) decreases from 110 at pristine condition to 52 only by a factor of 2 after decompression, indicating the remaining good sample quality. RRR of the pristine sample \#2 is  around 360, and it is greatly reduced to 35 by a factor of 10 after decompression.  The above results are consistent with the electron micrographs collected on these two samples, as shown in Fig.~S~3 \cite{CeAgSbSI}. The surface structure of sample \#1 after decompression remains intact. By comparison, sample \#2 shows signs of shallow cracks after decompression from a slightly higher stress, indicating considerable sources of scattering for electrical transport.

To the best of our knowledge, such stress-tuning studies on ferromagnetic coercivity have never been reported so far.  Intriguingly, the stress effect on the enhancement of coercivity is enormous, and the enhancement remains after decompression (Fig.~\ref{PD}). Prior to discussion of the stress-tuning of an extrinsic property (coercivity $H_{\rm{C}}$), we would like to address the intrinsic part (saturation moment $M_{\rm{sat}}$) under stress. The linear magnetostriction (1/$L_{0}$)d$(\Delta L)$/d$(H)$ along the crystal $c$-axis of CeAgSb$_2$ with field applied along the same axis, measured at 5~K, displays a positive value below 1~T \cite{02AdoarX}. Utilizing Maxwell's relation, $\lambda = -\mu_{0}$$\rho$(d$M$/d$p$), where $\lambda$ is the volume magnetostriction and $\rho$ is the mass density, magnetization should decrease with uniaxial stress applied along the $c$-axis in CeAgSb$_2$, qualitatively in line with the experimental result.

Stress can alter the single-ion anisotropy by generating a perturbation of crystal electric field as a result of compression and expansion of different crystalline directions. Such effects can modify  $H_{\rm{A}}$, thus possibly increase coercivity. However, this process should be reversible upon compression and decompression, which does not appear to be the case for our study. Furthermore, such small stress of 1~kbar only changes the lattice parameter by about 0.1~\% if we take a typical Young's modulus of 100~GPa, which is likely a negligible effect for tuning $H_{\rm{A}}$. Thus, the enhancement of coercivity due to change of crystal electric field is rendered unlikely.   
 
Therefore, the enhancement of $H_{\rm{C}}$ under stress should be addressed in terms of defect-induced domain wall pinning. The remaining high coercivity after decompression (irreversibility) corroborates such argument. As shown in Figs.~\ref{MT} (d) and (e), pristine sample \#1 displays a lower RRR of 110 compared with that of \#2 of 360, indicating a higher defect concentration in sample \#1. However, their coercivity $H_{\rm{C}}$ is of similar size (28~Oe for sample \#1 and 32~Oe for \#2), suggesting the defects (which are likely point defects due to a slightly different crystal growth conditions) prior to the application of stress are distinctly different from the ones that cause the large enhancement of the coercivity.    

The residual resistivity $\rho_{\rm{0}}$ of metals is caused by defect scattering. For our pristine CeAgSb$_2$ samples, $\rho_{\rm{0}}$ is around 0.5~$\mu\Omega$ cm, as shown in Fig.~S~4 \cite{CeAgSbSI}. Upon compression, $\rho_{\rm{0}}$ increases ascribed to stress-induced defects. Unfortunately, due to the special geometry of samples prepared for this stress study, a quantitative assessment of $\rho_{\rm{0}}$ of these samples is not accessible. Nevertheless, the deviation of normalized resistance in the ferromagnetic state between pristine samples and decompressed ones [Figs.~\ref{MT} (d) and (e)] can be reasoned with the defect-induced increase of $\rho_{\rm{0}}$. An alternative explanation is given by an enhanced spin-dependent scattering caused by stress-induced defects in combination with magnetic domain walls \cite{97PRLTat} (that are potentially strongly modified when compared to the pristine samples). This suggests a deviation of resistivity only in the ferromagnetic state consistent with our experimental findings.                       

When stress is applied to a crystalline metal, typically a certain type of one-dimensional defects -- dislocations -- forms \cite{67ActFel}. It is plausible that such dislocations act as the pinning centers impeding the domain wall movement, leading to an increase of coercivity. Another type of defects can occur under stress in CeAgSb$_2$ is two-dimensional stacking faults, owning to the quasi-2D crystal structure of CeAgSb$_2$. Stacking faults have been shown to harden cobalt \cite{15AEMAkd} and Co-Sm thin films \cite{15AEMAkd}. At  low concentration of stacking faults, the average distance between these faults is larger than the domain wall thickness, which is argued to be the ideal case for pinning the domain walls. At high stress, numerous stacking faults occur, which can destroy the integrity of the crystal structure and weaken the ferromagnetic exchange, thus the coercivity will decrease \cite{03PRBSor}. Such picture agrees with our observation of decreasing magnetization and coercivity for $p >$ 1~kbar and also for the ground samples, which contain a large amount of defects.

An intriguing question is how the unexpected rectangular tubular patterns of magnetic domains evolves under stress \cite{22JPCMRus}. Studying how stress modifies these topological patterns of domains could provide insights into the above question. On the other hand, stress-induced defects in CeAgSb$_2$ may alter the topology of domains, generating certain flavors of domain pattern which are ``hard'' to reverse in a magnetic field, giving rise to a large magnetic coercivity. 
  
We would like to emphasize that stress-tuning magnetic hysteresis has  been rarely employed in ferromagnets. Established methods to increase coercivity mainly consider composition-tuning and sintering/annealing temperature-tuning. As far as we are aware, our study sets the first reported example for the observation of a strong enhancement of coercivity utilizing stress, tuning a soft magnet into a hard one. And this approach can be almost non-destructive to other properties of the sample. It is also tempting to propose that stress here generates certain unique patterns of defects, which are not observed through other existing methods, yet, efficient in pinning domain walls. Equally importantly, our study provides a perfect platform with highly tunable coercivity, but in the forms of simpler and cleaner single crystals compared to previous studies which usually involve complex chemical compositions, intricate intergranular phases, and different flavors of defects. These advantages of our system provide an easier path to elucidate the mechanism of defect-enhanced coercivity. In essence, our approach can potentially be extended to other magnets and will trigger a series of related future studies.  To capture the nature and role of stress here, further investigations, such as magneto-optical Kerr imaging, scanning-electron microscopy, and electron holography, are highly needed.

In summary, we found a strong enhancement of coercivity induced by uniaxial stress in the Kondo lattice CeAgSb$_2$. We postulate that stress generates some forms of defects, that are highly efficient in pinning domain walls in CeAgSb$_2$.         

Experimental data associated with this manuscript are available from Ref.~\cite{DATA}.
\section{Acknowledgments}
 This work was funded by the Deutsche Forschungsgemeinschaft (DFG, German Research Foundation) -- TRR 360 -- 492547816. B.S.  acknowledges the financial support of Alexander von Humboldt Foundation.

\widetext
\clearpage
\begin{center}
	\textbf{\large Supplementary Material for \\ Strong enhancement of magnetic coercivity induced by uniaxial stress}
\end{center}
\setcounter{equation}{0}
\setcounter{figure}{0}
\setcounter{table}{0}
\setcounter{page}{1}
\makeatletter
\setcounter{section}{0}
\renewcommand{\thesection}{S-\Roman{section}}
\renewcommand{\thetable}{S\arabic{table}}
\renewcommand{\theequation}{S\arabic{equation}}
\renewcommand{\thefigure}{S\arabic{figure}}
\renewcommand{\bibnumfmt}[1]{[S#1]}
\renewcommand{\citenumfont}[1]{S#1}

\author{Bin Shen}
\affiliation  {Experimental Physics VI, Center for Electronic Correlations and Magnetism, University of Augsburg, 86159 Augsburg, Germany}

\author{Franziska Breitner}
\affiliation  {Experimental Physics VI, Center for Electronic Correlations and Magnetism, University of Augsburg, 86159 Augsburg, Germany}

\author{Philipp Gegenwart}
\affiliation{Experimental Physics VI, Center for Electronic Correlations and Magnetism, University of Augsburg, 86159 Augsburg, Germany}

\author{Anton Jesche}
\affiliation  {Experimental Physics VI, Center for Electronic Correlations and Magnetism, University of Augsburg, 86159 Augsburg, Germany}

\date{\today}


\maketitle


\section{Materials and Methods}

\subsection{Sample preparation}

Three single crystalline samples of CeAgSb$_2$ (\#1, \#2, and \#3) with plane-parallel faces were cut using diamond wire saw (IDRAS) into a dimension of around 1$\times$1$\times$0.5~mm$^3$, and faces perpendicular to the force direction were carefully polished with fine sandpapers. Side faces of sample \#1 were surrounded by Stycast 1266 to provide certain protection. 

\subsection{Magnetization measurement}
Magnetization along the $c$-axis under uniaxial stress along the same direction was collected in a Quantum Design MPMS-5S system using a commercial uniaxial pressure cell (XPC-5, Quantum Design). Uniaxial pressure was applied at room temperature. To obtain the magnetization of the sample, raw extraction data of the empty cell was first measured as the background, then subtracted from the signal containing inserted sample and the cell collected with the same procedure. The stress $p$ was calculated using the formula $p$ = $F/S$, where $F$ is the force applied to the sample and $S$ is the cross section area of the sample.

\section{Experimental results}

\subsection{Magnetization vs. magnetic field}
\begin{figure}
	\includegraphics[angle=0,width=0.45\textwidth]{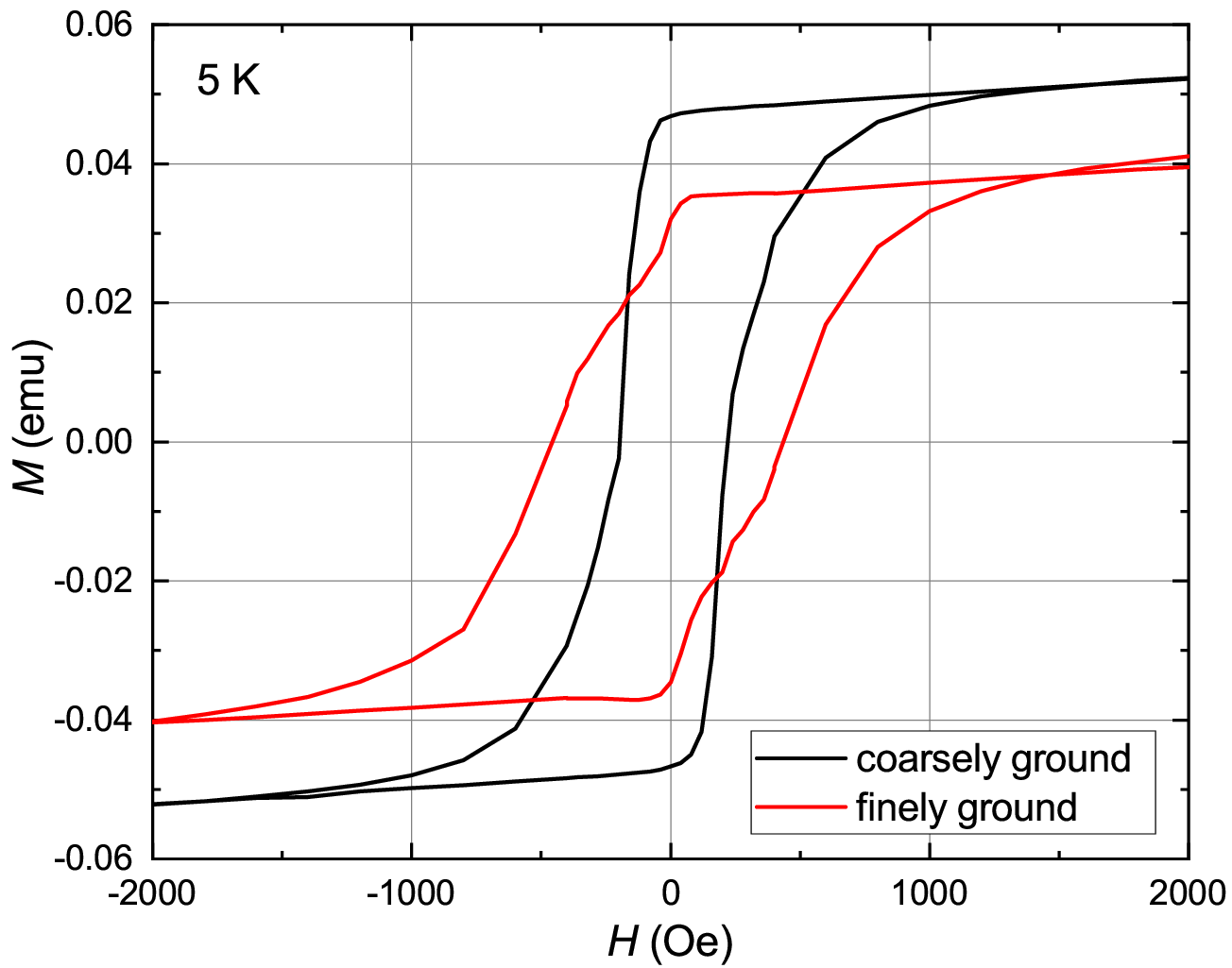}
	\vspace{-12pt} \caption{\label{MHGround}  $M(H)$ for coarsely and finely ground CeAgSb$_2$ crystals at 5~K.}
	\vspace{-12pt}
\end{figure}

Isothermal magnetization $M(H)$ was measured for coarsely and finely ground CeAgSb$_2$ samples, as shown in Fig.~\ref{MHGround}.

\subsection{Magnetization vs. temperature}

\begin{figure}
	\includegraphics[angle=0,width=0.4\textwidth]{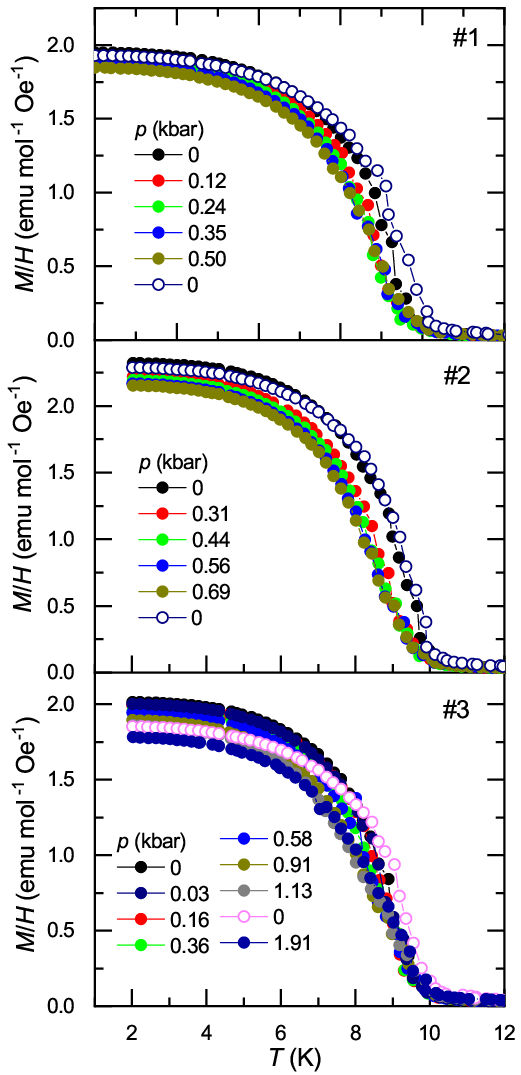}
	\vspace{-12pt} \caption{\label{MTSM} Temperature dependence of magnetization $M(T)$ at various stresses at a magnetic field of 0.1~T for sample \#1, \#2, and \#3. The open symbols were measured after decompression.}
	\vspace{-12pt}
\end{figure}

Temperature dependence of magnetization $M(T)$ at various stresses were measured for all the samples, as shown in Fig.~\ref{MTSM}.

\subsection{Electron micrographs}

\begin{figure}
	\includegraphics[angle=0,width=0.7\textwidth]{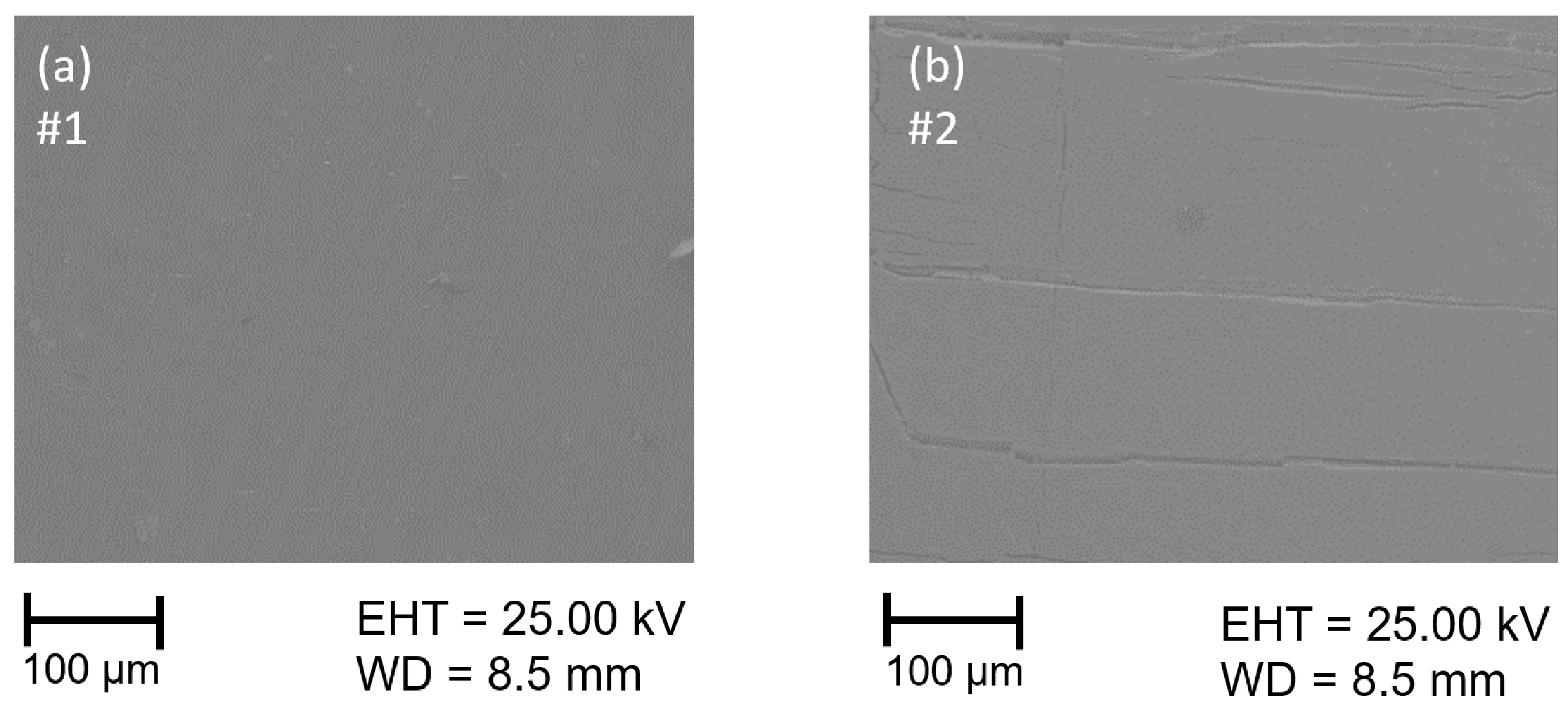}
	\vspace{-12pt} \caption{\label{Electron} Electron micrographs of (a) sample \#1 and (b) \#2 after decompression, respectively.}
	\vspace{-12pt}
\end{figure}

Electron micrographs were collected so as to examine the surface structure of sample \#1 and \#2 after decompression, as shown in Fig.~\ref{Electron}. Sample \#3 gets fragmented after decompression from a large stress of 1.91~kbar, manifesting the existence of severe cracks in the sample.              

\subsection{Resistivity}

\begin{figure}
	\includegraphics[angle=0,width=0.35\textwidth]{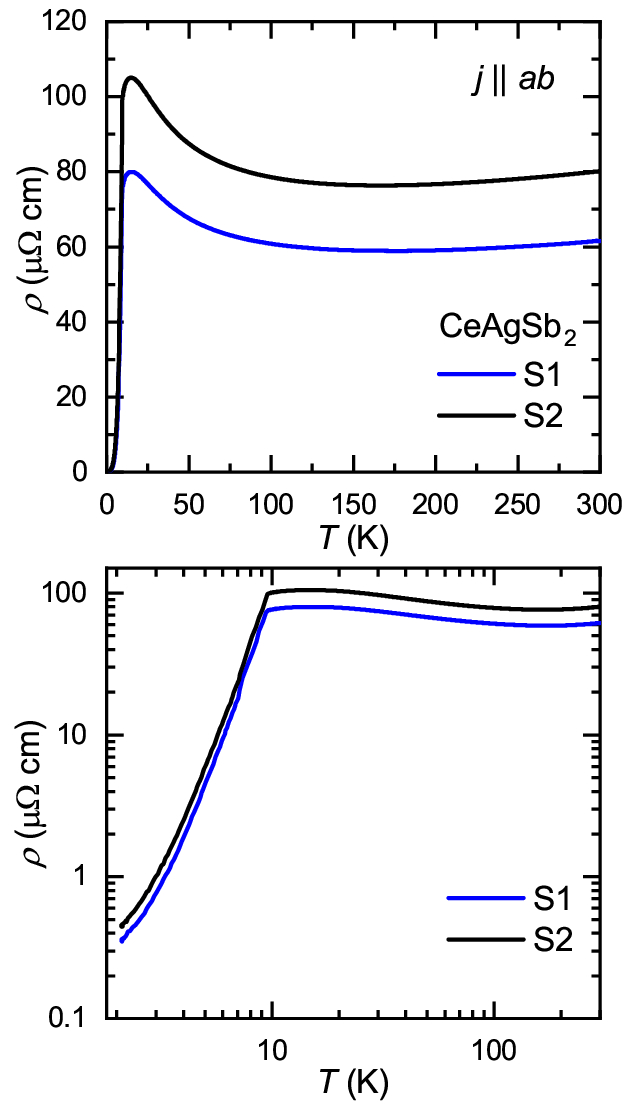}
	\vspace{-12pt} \caption{\label{RT_pri} Temperature dependence of resistivity $\rho(T)$ of two pristine CeAgSb$_2$ samples in two different presentations, with current applied along the $ab$-plane.}
	\vspace{-12pt}
\end{figure}

Figure~\ref{RT_pri} shows the resistivity $\rho(T)$ of two pristine CeAgSb$_2$ samples.

\end{document}